\documentclass[fleqn,usenatbib]{mnras}

\usepackage{newtxtext,newtxmath}

\newcommand{\rtwom}{R$_{200\textrm{m}}$}
\newcommand{\mtwom}{M$_{200\textrm{m}}$}
\newcommand{\rfivec}{R$_{500\textrm{c}}$}
\newcommand{\rtwofivec}{R$_{2500\textrm{c}}$}
\newcommand{\rmax}{R$_{\textrm{max}}$}
\newcommand{\pth}{P$_{\textrm{th}}$}
\newcommand{\pnt}{P$_{\textrm{nt}}$}
\newcommand{\ptot}{P$_{\textrm{tot}}$}

\usepackage[T1]{fontenc}

\DeclareRobustCommand{\VAN}[3]{#2}
\let\VANthebibliography\thebibliography
\def\thebibliography{\DeclareRobustCommand{\VAN}[3]{##3}\VANthebibliography}

\usepackage{graphicx}	
\usepackage{amsmath}	
\usepackage{mathtools}

\title[CLUMP-3D: Non-Thermal Motions]{CLUMP-3D: the Lack of Non-Thermal Motions in Galaxy Cluster Cores}

\author[J. Sayers et al.]
{Jack Sayers,$^{1}$\thanks{E-mail: jack@caltech.edu}
Mauro Sereno,$^{2,3}$
Stefano Ettori,$^{2,3}$
Elena Rasia,$^{4}$
Weiguang Cui,$^{5}$
Sunil Golwala,$^{1}$
\newauthor
Keiichi Umetsu,$^{6}$
and Gustavo Yepes$^{7,8}$
\\
$^{1}$California Institute of Technology, 1200 East California Boulevard, Pasadena, California 91125, USA\\
$^{2}$Osservatorio di Astrofisica e Scienza dello Spazio di Bologna, via Piero Gobetti 93/3, I-40129 Bologna, Italy\\
$^{3}$INFN, Sezione di Bologna, viale Berti Pichat 6/2, 40127 Bologna, Italy\\
$^{4}$INAF - Osservatorio Astronomico di Trieste, via Tiepolo 11, I-34143 Trieste, Italy\\
$^{5}$Institute for Astronomy, University of Edinburgh, Royal Observatory, Edinburgh EH9 3HJ, United Kingdom\\
$^{6}$Academia Sinica Institute of Astronomy and Astrophysics (ASIAA), No. 1, Section 4, Roosevelt Road, Taipei 10617, Taiwan\\
$^{7}$Departamento de F\'{\i}sica Te\'orica M-8, Universidad Aut\'onoma de Madrid, Cantoblanco, E-28049 Madrid, Spain\\
$^{8}$Centro de Investigaci\'on Avanzada en F\'{\i}sica Fundamental (CIAFF), Universidad Aut\'onoma de Madrid, 28049 Madrid, Spain
}

\date{Accepted XXX. Received YYY; in original form ZZZ}

\pubyear{2021}

\begin{document}
\label{firstpage}
\pagerange{\pageref{firstpage}--\pageref{lastpage}}
\maketitle

\begin{abstract}
  We report the non-thermal pressure fraction (\pnt/\ptot)
  obtained from a three-dimensional triaxial analysis
  of 16 galaxy clusters in the CLASH sample
  using gravitational lensing (GL) data primarily from {\it Subaru} and {\it HST},
  X-ray spectroscopic imaging from {\it Chandra}, and Sunyaev-Zel'dovich effect (SZE)
  data from {\it Planck} and Bolocam.
  Our results span the approximate radial range 0.015--0.4\rtwom\ ($\sim 35$--1000 kpc).
  At cluster-centric radii smaller than
  0.1\rtwom\ the ensemble average \pnt/\ptot\
  is consistent with zero with an upper limit of nine per cent,
  indicating that heating from active galactic nuclei and other relevant
  processes does not produce significant deviations from hydrostatic equilibrium (HSE).
  The ensemble average \pnt/\ptot\ increases outside of this radius
  to approximately 20 per cent at 0.4\rtwom,
  as expected from simulations, due to newly accreted material
  thermalizing via a series of shocks.
  Also in agreement with simulations, we find significant cluster-to-cluster
  variation in \pnt/\ptot\ and little difference in the ensemble average
  \pnt/\ptot\ based on dynamical state.
  We conclude that on average, even for diverse samples,
  HSE-derived masses in the very central regions of galaxy clusters
  require only modest corrections due to non-thermal
  motions.
\end{abstract}




\section{Introduction} \label{sec:intro}

Given the hierarchical buildup of structure in the Universe,
galaxy clusters are the largest and most recent objects to
form \citep{Davis1985, Kravtsov2012}.
Their evolution is dictated mainly by gravity \citep{Kaiser1986,Kaiser1991},
but a range of more complicated
and less understood processes also play a significant
role \citep[e.g.,][]{McNamara2007,Markevitch2007,Brunetti2007}.
As a result, the diffuse ionized gas of the intra-cluster medium (ICM),
which contains most of the baryons, does not reside in strict hydrostatic
equilibrium (HSE) with the gravitational
potential \citep{Pratt2019,Ansarifard2020}.

For instance, radiative cooling is capable of producing large
reservoirs of cold star-forming gas in the central regions,
but such reservoirs are not observed \citep{Peterson2006}.
While the cooling may be offset and/or suppressed by a range of processes,
including core sloshing \citep{Markevitch2007},
dynamical friction from galaxy motions \citep{El-Zant2004},
and conduction and turbulent mixing \citep{Ruszkowski2010},
feedback from centrally located active galactic nuclei (AGN) is thought to be the
primary heating source \citep{McNamara2007,Gitti2012}.
This feedback is dynamic: AGN jets inflate
bubbles in the gas and transfer energy via multiple mechanisms,
such as shocks, cavity heating, and convective mixing \citep{Yang2016}.

Recently, {\it Hitomi} used high resolution X-ray spectroscopy
to measure the gas velocity structure of Perseus's
core.
Surprisingly, despite its active AGN and the presence of features like
a sloshing cold front \citep{Simionescu2012},
the gas was found to be remarkably quiescent, with
a non-thermal pressure fraction \pnt/\ptot~$\simeq$~4 per cent \citep{Hitomi2016}.
More recently, a multi-probe
observational study of five galaxy clusters at $z \simeq 0.35$ with extremely round
morphologies from the Joint Analysis of Cluster Observations
(JACO) project found \pnt/\ptot~$\lesssim 6$ per cent within \rtwofivec\
\citep[corresponding to $\sim 0.2$\rtwom,][]{Siegel2018}.\footnote{  
  We refer to radii, and corresponding masses, that
  enclose an average density defined by the subscript. The number in the subscript
  denotes the overdensity relative to the background critical (c) or
  matter (m) density of the universe.
  All physical values described in this work
  were computed assuming a flat $\Lambda$CDM cosmology with
  $\Omega_{\textrm{m}} = 0.3$, $\Omega_{\Lambda} = 0.7$, and $h=0.7$.}
This suggests that
Perseus is not unusual, although the strict selection
of the JACO sample may not be representative of the overall
population.

External to these central regions, simulations predict that \pnt/\ptot\
increases with radius as a result of the series of shocks required
to thermalize newly accreted material \citep{Miniati2000,Molnar2012,Nelson2014,Shi2014,Lau2015,Shi2015}.
Recently, JACO \citep{Siegel2018} and
the XMM-{\it Newton} Cluster Outskirts Project \citep[X-COP,][]{Eckert2019}, which included
12 local, mostly relaxed, massive galaxy clusters, both found
\pnt/\ptot~$\lesssim 10$ per cent
at \rfivec\ ($\sim 0.5$\rtwom). These results are in some tension with simulation-based
predictions of 15--25 per cent at this radius \citep{Nelson2014,Angelinelli2020,Gianfagna2020},
and potentially imply more efficient thermalization.

The result presented here is obtained from modeling multi-probe
observations using the CLUster Multi Probe - 3 Dimensions (CLUMP-3D)
package \citep{Sereno2017}.
The available multi-wavelength data enable independent measurements
of the thermal pressure (\pth) and total mass distribution
(which sets the total pressure, \ptot, required to offset gravity).
The non-thermal pressure (\pnt) is the difference
between \ptot\ and \pth.
While this basic formalism is common to CLUMP-3D, JACO, and X-COP,
our model relies on fewer assumptions.
First and foremost, we do not use a spherically symmetric model,
which can significantly bias the results from multi-probe reconstructions
due to the strong degeneracy between the line of sight extent
of the galaxy cluster and the inferred non-thermal pressure
fraction \citep[e.g., see the discussion in Section 7.2 of][]{Siegel2018}.
In addition, we also allow for intrinsic cluster-to-cluster
scatter in \pnt/\ptot.
Unlike JACO, we do not assume a fixed radial
form for \pnt/\ptot.
Unlike X-COP, we measure the total mass with gravitational lensing (GL).
Like JACO, we use strong lensing (SL) constraints
to reliably probe mass profiles near the core.
In sum, our technique 
enables a more flexible, data-driven approach that can address larger and more
diverse samples.

\section{Data} \label{sec:data}

Because of the deep multi-probe observations required for this type of analysis,
previous applications of similar methods have been limited to
individual galaxy clusters
\citep[e.g.,][]{Morandi2012,Limousin2013,Sereno2013}.
In this work, we use the datasets
obtained as part of the Multi-Cycle Treasury program
Cluster Lensing and Supernova survey with Hubble \citep[CLASH,][]{Postman2012}
to model 16 individual objects.
While CLASH includes 25 galaxy clusters, five
were selected based on lensing strength and appear to be dynamically complicated systems
that may not be accurately modeled within the CLUMP-3D formalism
\citep[i.e., at least four of these five objects are undergoing major mergers between
at least two distinct sub-clusters, see][]{Mann2012,Postman2012}.
Of the remaining 20, four lack the requisite ground-based wide-field
GL data required for our analysis \citep{Umetsu2018}, leaving a sample of 16 for this
work (see Table~\ref{tab:sample}).

All of these clusters have a regular X-ray morphology, 
which indicates a higher than average probability of being
dynamically relaxed \citep{Meneghetti2014}.
None of the galaxy clusters appear to be undergoing a major merger.
However, eight of the 16 show potential signs of some merger activity
in at least one systematic search for such objects based on X-ray imaging,
location of the central galaxy, and/or member-galaxy velocity
dispersions \citep{Gilmour2009,Postman2012,Mann2012}.

\begin{table}
  \centering
  \caption{Galaxy cluster sample. \rtwom\ (and \mtwom)
    corresponds to the spherical volume determined from
    our fits to the CLUMP-3D model \citep{Sereno2018}. \rmax\ is the maximum
    radius probed by the {\it Chandra} observations.
    While all of the objects have a regular
    X-ray morphology, which suggests they are more likely than average to be dynamically relaxed,
    those with any evidence of potential merger activity are indicated by
    stars \citep{Gilmour2009,Postman2012,Mann2012,Meneghetti2014}.}
  \label{tab:sample}
  \begin{tabular}{lcccc}
    Cluster Name & Redshift & \mtwom & \rtwom & \rmax \\
    & & $10^{14}$ M$_{\sun}$ & Mpc & \rtwom \\ \hline
    ABELL 0383 & 0.188 & $\phantom{1}8.0 \pm 1.6$ & $2.41 \pm 0.16$ & 0.39 \\
    ABELL 0209$^*$ & 0.206 & $11.4 \pm 4.0$ & $2.67 \pm 0.31$ & 0.51 \\
    ABELL 2261$^*$ & 0.225 & $23.3 \pm 4.0$ & $3.36 \pm 0.19$ & 0.32 \\
    RX J2129.6+0005$^*$ & 0.234 & $\phantom{1}7.0 \pm 1.9$ & $2.22 \pm 0.20$ & 0.62 \\
    ABELL 0611 & 0.288 & $\phantom{1}8.9 \pm 2.1$ & $2.30 \pm 0.19$ & 0.31 \\
    MACS J2140.2-2339 & 0.313 & $\phantom{1}9.4 \pm 2.9$ & $2.30 \pm 0.19$ & 0.22 \\
    ABELL S1063$^*$ & 0.348 & $16.6 \pm 3.7$ & $2.71 \pm 0.20$ & 0.50 \\
    MACS J1115.8+0129 & 0.352 & $18.9 \pm 3.9$ & $2.82 \pm 0.19$ & 0.55 \\
    MACS J1931.8-2635 & 0.352 & $\phantom{1}7.6 \pm 2.0$ & $2.08 \pm 0.18$ & 0.68 \\
    MACS J1532.8+3021 & 0.363 & $\phantom{1}7.3 \pm 1.9$ & $2.03 \pm 0.17$ & 0.39 \\
    MACS J1720.2+3536 & 0.391 & $10.4 \pm 2.4$ & $2.25 \pm 0.17$ & 0.45 \\
    MACS J0429.6-0253 & 0.399 & $\phantom{1}7.3 \pm 1.4$ & $1.98 \pm 0.13$ & 0.27 \\
    MACS J1206.2-0847$^*$ & 0.440 & $17.1 \pm 2.1$ & $2.56 \pm 0.11$ & 0.41 \\
    MACS J0329.6-0211$^*$ & 0.450 & $11.9 \pm 2.9$ & $2.25 \pm 0.14$ & 0.28 \\
    MACS J1347.5-1144$^*$ & 0.451 & $35.0 \pm 6.6$ & $3.23 \pm 0.20$ & 0.60 \\
    MACS J0744.9+3927$^*$ & 0.686 & $17.6 \pm 5.4$ & $2.21 \pm 0.28$ & 0.54 
  \end{tabular}
\end{table}

Observational data available for the full sample include:
X-ray spectroscopic imaging from
{\it Chandra} with a median exposure time of 44~ksec \citep{Sereno2018};
Sunyaev-Zel'dovich effect (SZE) imaging from Bolocam and
{\it Planck} with a median combined signal-to-noise ratio of 12.3 \citep{Sayers2016};
wide-field ground-based weak lensing (WL) constraints from an average
background source density of 12~arcmin$^{-2}$ after stringent color--color cuts
\citep[mainly using
{\it Subaru} imaging from $\ge 3$ bands,][]{Umetsu2014,Umetsu2018};
and 16-filter {\it HST} imaging providing a median
of 18 effective SL constraints per galaxy cluster and 
an average background source density of 50~arcmin$^{-2}$ for WL \citep{Zitrin2015}.

\section{Methods} \label{sec:methods}

\begin{figure*}
  \centering \includegraphics[width=0.85\textwidth]{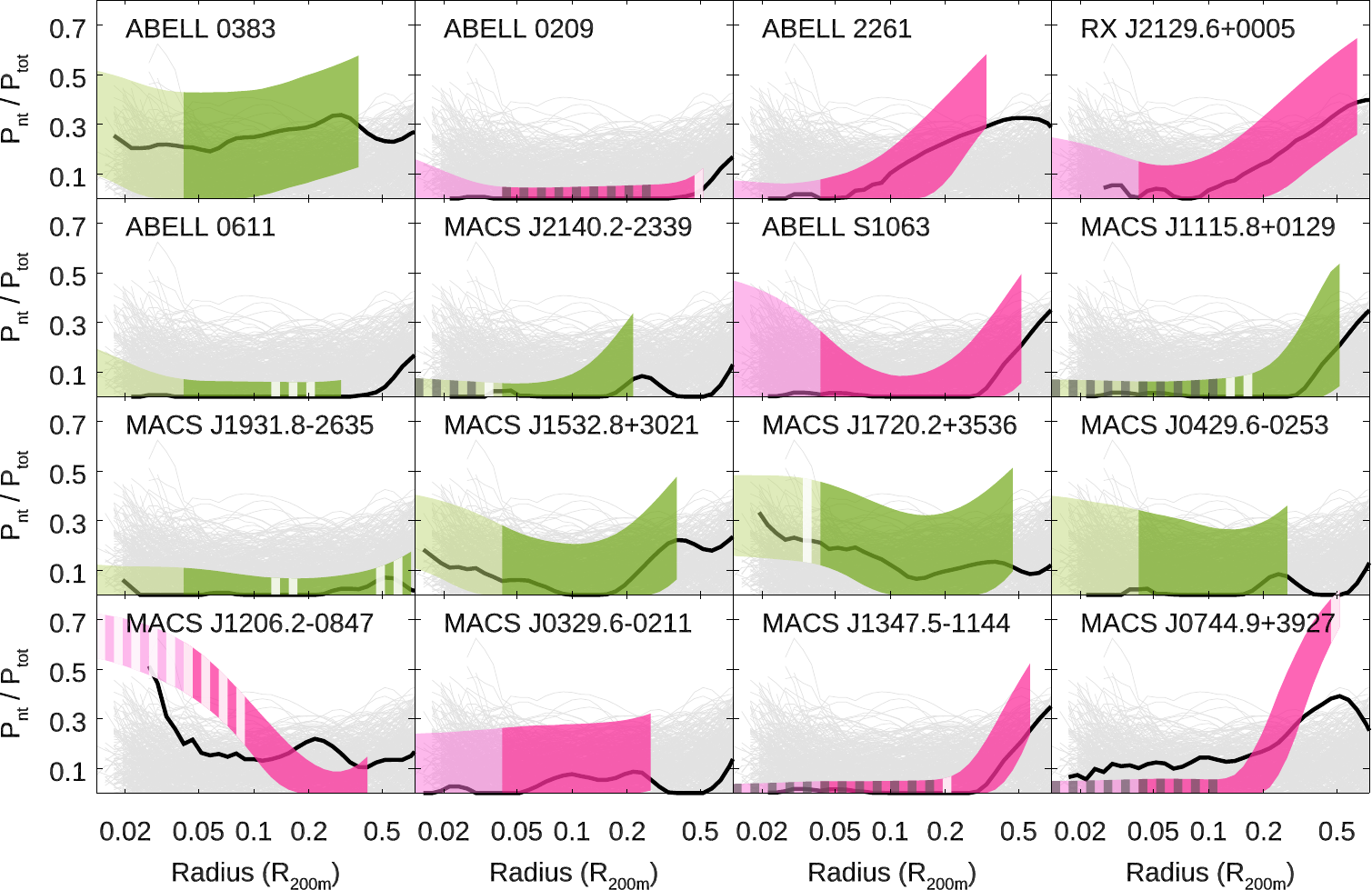}
  \caption{The shaded bands indicate
    95 per cent confidence regions of \pnt/\ptot\ for each galaxy cluster,
    restricted to physically allowed values $\ge 0$
    using the technique of \citet{Feldman1998}.
    Pink denotes objects with potential merger activity and green denotes non-mergers
    \citep{Gilmour2009,Postman2012,Mann2012,Meneghetti2014}.
    Potential systematic errors 
    in our modeling may result in unaccounted for biases at $\textrm{R} \le 0.04$\rtwom,
    which is denoted with a lighter color.
    Dark grey hashes indicate values of \pnt/\ptot\ removed from the
    analysis based on goodness-of-fit to the physically allowed region and
    white hashes indicate values of \pnt/\ptot\ removed from
    the analysis based on goodness-of-fit to the ensemble average profile
    (see text for additional details).
    The profiles for 315 galaxy clusters from the 300 simulations \citep{Cui2018}
    are shown as thin grey lines, with the best match to the observed profile highlighted
    as a thicker black line.} 
  \label{fig:two}
\end{figure*}

\subsection{Three-Dimensional Model}

The CLUMP-3D model assumes an elliptical triaxial geometry,
with co-alignment and co-centering of the major, intermediate, and minor axes of both
the total mass and gas distributions. The eccentricities 
are allowed to separately vary, with the well-motivated prior that the gas
distribution is rounder \citep{Lau2011}.
As a function of the elliptical radial coordinate,
the total mass density is parameterized by
the Navarro-Frenk-White (NFW) profile \citep{Navarro1996},
while the gas density and temperature are parameterized by a modified beta-model
and a modified broken power law \citep{Vikhlinin2006}.
The model has seven free parameters related to the shape
and orientation of the galaxy cluster: two axial ratios for both the total
mass density and the gas density, along with three angles to describe the
orientation in the observer's reference frame.
In addition, the total mass density has two free
parameters, the gas density has six free parameters, and the
gas temperature has five free parameters.

Within this model, X-ray and SZE data mostly constrain the parameters related
to the gas, while
the GL data constrain the parameters related to the
total mass.
Specifically, the X-ray surface brightness is proportional to
$\int \rho_{\textrm{gas}}^2 \Lambda dl$,
where $\rho_{\textrm{gas}}$ is the gas density,
$\Lambda$ is the
X-ray cooling function, and $dl$ is along the line of sight through the galaxy
cluster.
Independently, the gas temperature ($\textrm{T}_{\textrm{gas}}$) can be determined
from X-ray spectroscopy.
The SZE brightness is proportional to
$\int \textrm{P}_{\textrm{th}} dl$,
where $\textrm{P}_{\textrm{th}} \propto \rho_{\textrm{gas}} \textrm{T}_{\textrm{gas}}$.
Because the X-ray and SZE data redundantly probe the value of \pth,
but with a different dependence on the integral $dl$, their
combination measures the line of sight extent
of the galaxy cluster and thus its three dimensional geometry.
The GL data directly probe the projected mass density, which can
then be used to measure the three dimensional total mass distribution
based on the geometry which is primarily determined from the X-ray and SZE data.

From the observational data available,
the following products are generated to constrain the model using the brightest
cluster galaxy as the centre \citep{Sereno2017,Sereno2018}.
Background and exposure-corrected {\it Chandra} surface brightness images are produced
in the 0.7--2.0~keV band using the CIAO 4.8 software and the
calibration database CALDB 4.7.1. Point sources are filtered out, and
the two-dimensional image is restricted to the circular region enclosing
80 per cent of the total source emission. Spectra used to constrain the gas
temperature are extracted from circular annuli and analyzed with
the XSPEC v.12.9 software. The SZE brightness is computed in circular
annuli using publicly available maps from
Bolocam \citep{Sayers2013} and {\it Planck} \citep{Planck2016} via a combined
analysis of both datasets \citep{Sayers2016}.
For the ground-based wide-field WL, two-dimensional
projected mass maps and their pixel--pixel covariance matrices are
obtained from a joint analysis of the shear and magnification bias
over a 24~arcmin~$\times$~24~arcmin square region \citep{Umetsu2018}.
Using the {\it Hubble Space Telescope} WL and SL data,
projected masses are computed in circular annuli between 5~arcsec and twice
the Einstein radius from the publicly available CLASH
PIEMDeNFW maps \citep{Zitrin2015}.

To compare with these products generated from the observational data,
the CLUMP-3D model is used to create projected GL
mass maps, X-ray surface brightness and temperature maps, and
SZE brightness maps from a given set of parameter values.
A Bayesian inference scheme is used to assess the probability distributions
of the parameters \citep{Sereno2017}. Priors spanning
large regions of parameter space, with uniform distributions in either
linear or logarithmic spaced intervals, are assumed. Some priors span
the full range of physically motivated or allowed values. For example,
the major axial ratio of the total mass density has a uniform linear prior
between 0.1 and 1.0 \citep{Jing2002}.
All other priors span a sufficiently large range to include any reasonable derived
value.

\subsection{Non-Thermal Pressure Analysis}

From these fits, we compute probability densities for the values
of $\rho_{\textrm{gas}}$ and the gravitational potential $\Phi_{\textrm{mat}}$
in three dimensions.
Here, we assume the gas to be in equilibrium, and so the ICM axial
ratios are set equal to the average ratios of the gravitational
potential.
We then determine \ptot\ from the HSE equation
($  \nabla \textrm{P}_{\textrm{tot}} = 
  -\rho_{\textrm{gas}} \nabla \Phi_{\textrm{mat}}$).
Using the three dimensional values for \pth\ and \ptot, we 
compute the average for each
within a set of discrete logarithmically spaced spherical annuli.
We find that the probability densities are well described by a Gaussian
distribution for $\ln($\pth/\ptot), and so profiles are generated
in this quantity for each galaxy cluster.
Finally, the radial coordinate is scaled according to
$\mathcal{R} = \textrm{R}$/\rtwom, which simulations indicate minimizes the cluster-to-cluster
scatter in \pnt/\ptot\ \citep{Nelson2014}.
Plots of \pnt/\ptot\ for all 16 galaxy clusters, which is computed
from the values of $\ln($\pth/\ptot) according to
\pnt~$=$~\ptot~$-$~\pth, are shown in
Figure~\ref{fig:two}.

From these radially scaled profiles, we constrain the ensemble-average
profile at each discrete radius using the log likelihood
\begin{equation}
  \mathcal{L}(X|\mu,\sigma_{\textrm{int}}) = 
  -\frac{1}{2} \sum_{i}^{N} \left[
    \frac{(x_i - \mu)^2}{\sigma_i^2 + \sigma_{\textrm{int}}^2} +
    \ln(\sigma_i^2 + \sigma_{\textrm{int}}^2) + \ln(2\pi) \right]
  \label{eqn:like}
\end{equation}
where $X_{\textrm{m}} = \{x_1,...,x_{N} \}$
are the values of $\ln($\pth/\ptot) with measurement
uncertainties $\sigma_i$ for the $N$ objects in our sample,
$\mu$ is the ensemble average,
and $\sigma_{\textrm{int}}$ is the intrinsic 
scatter about the average. This likelihood is evaluated using
a grid of $\mu \in [-2,2]$ and $\sigma_{\textrm{int}} \in [0,1]$
to obtain probability distributions for $\mu$
and $\sigma_{\textrm{int}}$. Note that the grid ranges are sufficiently
large to include all values of $\mu$ and $\sigma_{\textrm{int}}^2$
with non-negligible likelihood values. 

To mitigate the impact of outlier values on our ensemble analysis,
we discard any $x_i$ that fails the goodness-of-fit
criteria suggested by \citet{Feldman1998}.
To ensure consistency with the physically allowed region based on this
procedure, any individual $x_i$ with
\begin{equation}
  \int_{x_i}^{\infty} \mathcal{L}(x|\mu\!=\!0, \sigma_{\textrm{int}}\!=\!0)dx < 0.01
  \label{eqn:goodness_of_fit}
\end{equation}
is removed from the analysis. This occurs at a subset of radii
for five galaxy clusters, primarily at $\mathcal{R} \lesssim 0.1$\rtwom\
(see Figures~\ref{fig:two} and \ref{fig:ln_pdf}). As detailed in Section~\ref{sec:discussion},
our assumption of coalignment between the gas density and the
gravitational potential may be invalid at $\mathcal{R} \lesssim 0.04$\rtwom.
We therefore suspect biases related to the breakdown of this assumption are the
underlying cause of the un-physical values seen in these five clusters,
which provides additional motivation for removing them
from the ensemble analysis.

To further minimize contamination from outliers,
we also remove any $x_i$ that
sufficiently differs from the maximum likelihood value of
$\mu$ such that they would be expected to occur less than one per cent
of the time given the maximum likelihood value of $\sigma_{\textrm{int}}$.
This removal is performed in an iterative manner until no such $x_i$ remain,
and primarily impacts one object, MACS J1206.2-0847, at
$\mathcal{R} \lesssim 0.1$\rtwom, where it has a much larger non-thermal pressure
fraction than any other galaxy cluster in our sample.
As with the un-physical \pnt/\ptot\ values noted above,
we suspect these values are also due to a breakdown of
our assumption of coalignment between the gas density
and the gravitational potential.
In addition to MACS J1206.2-0847,
data are also removed from other objects based on this criteria.
However, in those instances the removal
is limited to a narrow radial range
(see Figures~\ref{fig:two} and \ref{fig:ln_pdf}).

From a total of 16 values, one for each galaxy cluster,
the number of $x_i$ eliminated at a given radius due to
the combination of both removal processes detailed above is between 1--6,
and so the majority of the sample is retained for the ensemble fits.
The value of $\mu$ is largely insensitive to these data cuts,\footnote{
  It may appear that the data cuts will shift the ensemble mean \pnt/\ptot\ to lower
  values at small radii due to the exclusion of data from MACS~J1206.2-0847
  (see Figure~\ref{fig:two}). However, in aggregate, the data cuts
  actually result in slightly higher values for the ensemble mean \pnt/\ptot\
  at those radii, since data are also excluded from other galaxy clusters
  with very low values of \pnt/\ptot.}
even if no $x_i$ are removed. However, the value of $\sigma_{\textrm{int}}$
does increase if fewer $x_i$ are removed, particularly at small radii where
it can be up to a factor of $\simeq 1.5$ larger when all the $x_i$ are retained.

\subsection{Confidence Intervals in the Physically Allowed Region}

We evaluate the likelihood at $\mu > 0$, which is
in the physically disallowed region corresponding to
\pth~$>$~\ptot\ (i.e., \pnt~$<$~0).
Establishing confidence intervals on the values of $\mu$ 
requires care due to this physical boundary. For instance,
imposing the physical boundary on
the $x_i$ prior to computing $\mu$ can result in
biases \citep{Leccardi2008}. Furthermore, \citet{Feldman1998}
have argued that simple prescriptions such as Bayesian priors
or renormalization of the Bayesian posterior can
produce intervals without the desired coverage.
For this analysis, we adopt the \cite{Feldman1998} approach for
establishing frequentist confidence intervals in the physically
allowed region.

\begin{figure}
  \centering \includegraphics[width=0.4\textwidth]{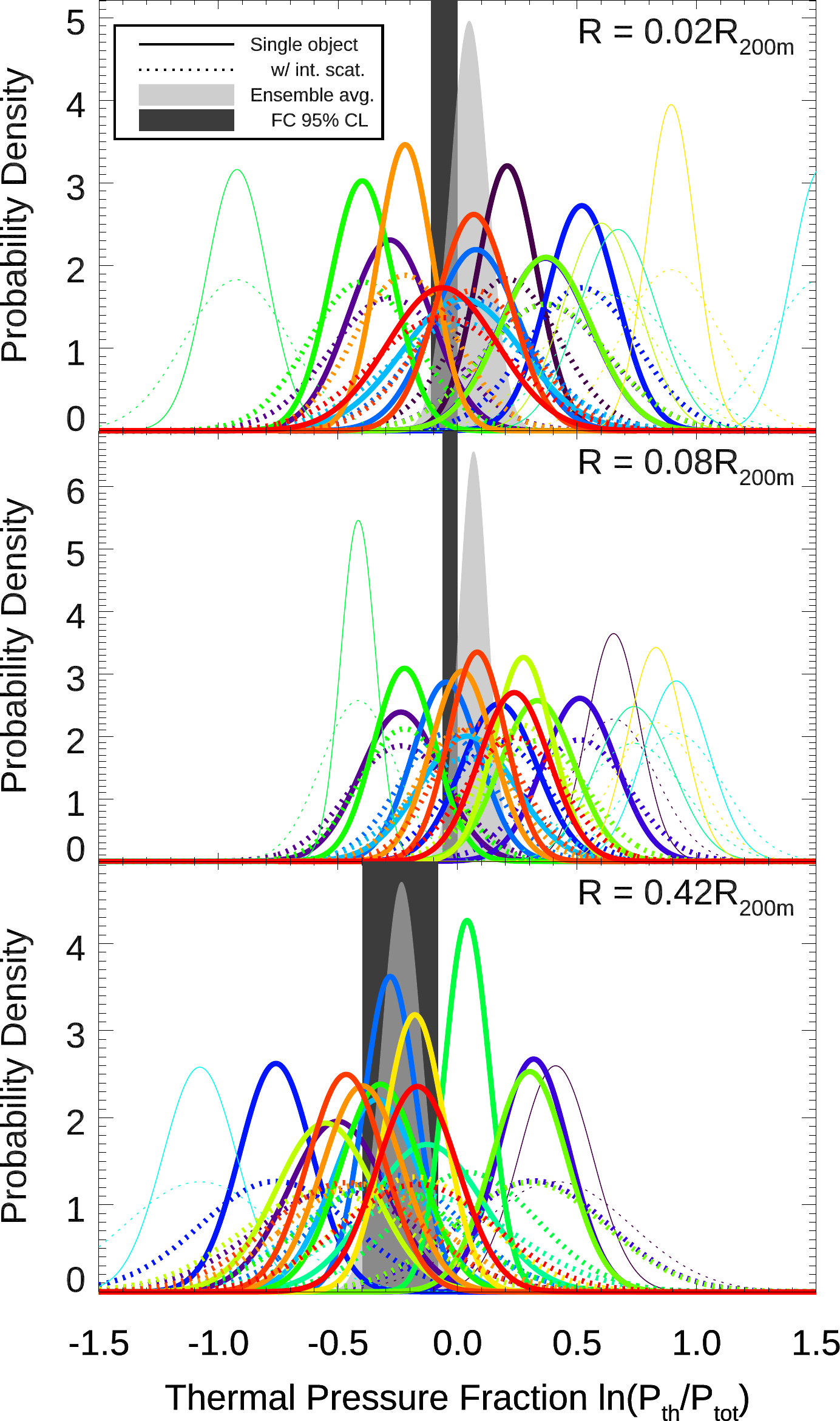}
  \caption{Measured probability densities for the logarithmic thermal
    pressure fraction ln(\pth/\ptot) at three different radii.
    Each solid line represents a single
    cluster when considering only the statistical uncertainties. The dashed
    lines indicate the effective broadening of the distributions when
    the fitted intrinsic scatter is included. The shaded grey region indicates
    the distribution for the ensemble average, including intrinsic scatter. Solid
    black encloses the 95 per cent confidence region within the physically allowed
    set of values based on the \citet{Feldman1998} technique
    (i.e., ln(\pth/\ptot)~$\le 0$).
    Thinner lines represent galaxy
    clusters that were excluded from the ensemble analysis based on the 
    criteria described in the text.}
  \label{fig:ln_pdf}
\end{figure}

Specifically, a value of $\mu$ should be included in the
$\alpha$ per cent confidence interval if the data $X_{\textrm{m}}$
are in the $\alpha$ per cent most likely observational
outcomes were $\mu$ the true underlying value.
As described by \citet{Feldman1998},
the statistic for evaluating whether an outcome is in the
$\alpha$ per cent most likely outcomes is
\begin{equation}
  R(X) = \frac{\mathcal{L}(X|\mu)}{\mathcal{L}(X|\mu^{\star})},
  \label{eqn:r}
\end{equation}
where $\mu^{\star}$ maximizes the likelihood within the physically
allowed region (i.e., $\mu^{\star} \le 0$) and $\mathcal{L}(X|\mu)$
is obtained from Equation~\ref{eqn:like}.
Absent a physical boundary, the denominator is independent of
$X$ and $R(X) \propto \mathcal{L}(X|\mu)$.
Because every potential observational outcome $X$ yields a value $R(X)$,
$\mathcal{L}(X|\mu)$ maps into a likelihood
$\mathcal{L}^{\prime}(R|\mu)$. We define a value $R_{\alpha}$ such
that the integral of $\mathcal{L}^{\prime}(R|\mu)$
over $R(X) \ge R_{\alpha}$ is equal to $\alpha$.
We include a value of $\mu$ in the $\alpha$ per cent confidence interval
if $R(X_{\textrm{m}}|\mu) \ge R_{\alpha}$.
To obtain the complete desired confidence interval, this process is
performed over a grid of values of $\mu$.

To assess the validity of our measured values of $\mu$ given the physical
boundary $\mu \le 0$,
we again consider the \citet{Feldman1998} goodness-of-fit
test described above and quantified in Equation~\ref{eqn:goodness_of_fit}
for the case of the individual $x_i$. We perform an analogous test for the
values of $\mu$ obtained in our grid search at each radius. The worst
goodness-of-fit values occur at intermediate radii from $\sim 0.05$--0.09\rtwom,
and correspond to a probability to exceed of approximately 0.70--0.80 (e.g., see
the middle panel of Figure~\ref{fig:ln_pdf}). This goodness-of-fit is
significantly better than the threshold suggested by \citet{Feldman1998},
corresponding to a probability to exceed of 0.99, and suggests
that none of our measured $\mu$ fall sufficiently outside
of the physically allowed region to be considered invalid.

\subsection{Analysis of Simulated Galaxy Clusters}

We apply the same analysis to \pth\ and \ptot\ profiles
obtained in spherical annuli
from 315 simulated galaxy clusters in the Three Hundred Project
\citep[hereafter ``300'',][]{Cui2018}.
This sample was selected based on the criteria described in \citet{Ansarifard2020},
and excludes nine objects with at least
one low-resolution particle, which are used to trace the large scale structure in
which the galaxy cluster is embedded, within \rtwom.
These halos span the approximate mass range \mtwom~$\simeq 10$--$25 \times 10^{14}$~M$_{\odot}$,
nearly identical to that of the CLASH objects,
and they were taken from a snapshot at $z=0.333$, close
to the median $z$ of $0.352$ for our observational sample.

\section{Results and Discussion} \label{sec:discussion}

\begin{figure}
  \centering
  \includegraphics[width=0.45\textwidth]{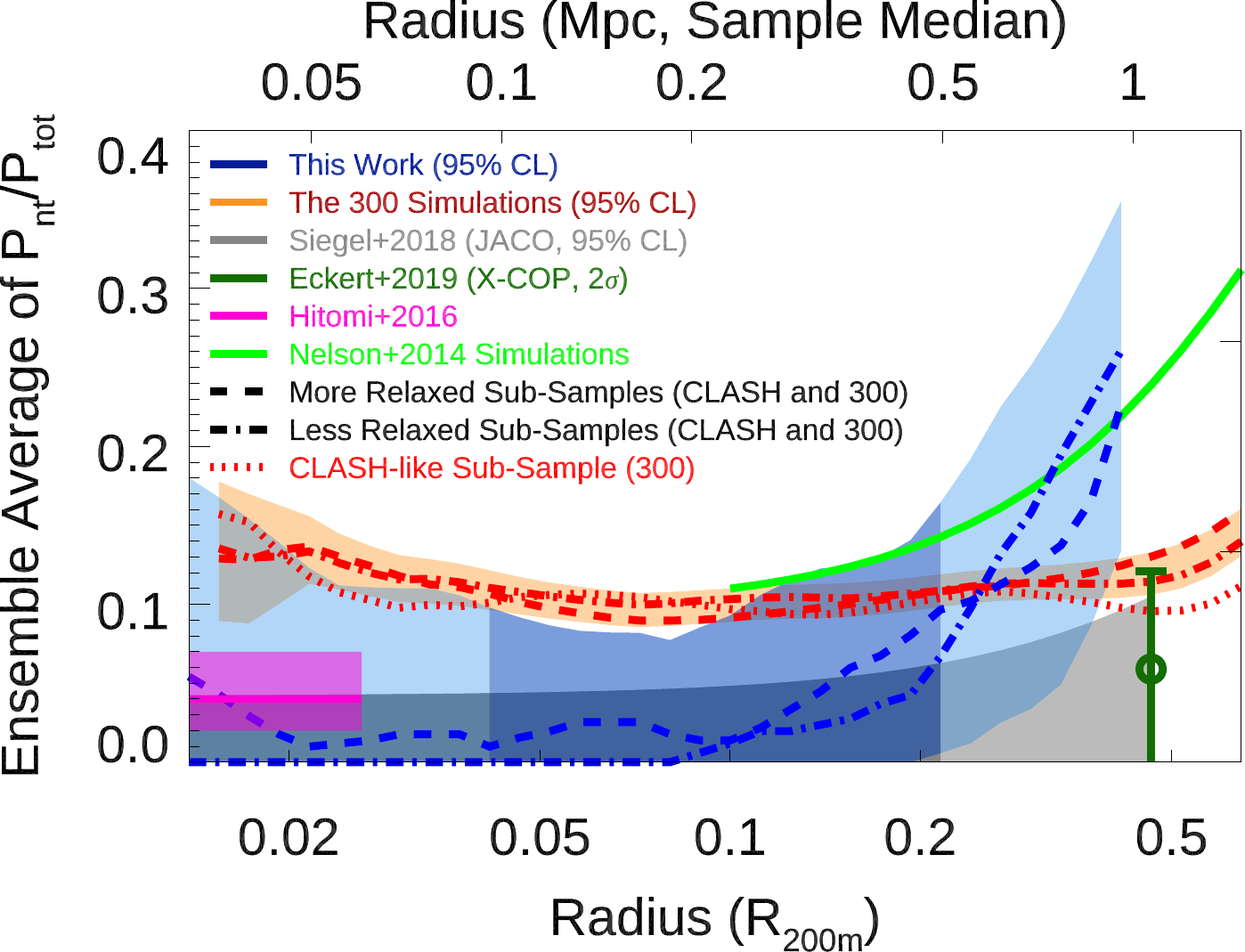}
  \caption{Ensemble-average \pnt/\ptot\
    from our analysis of 16 CLASH galaxy clusters (blue 95 per cent
    confidence region obtained from the \citet{Feldman1998} procedure
    detailed in the text and corresponding to the solid black regions
    in Figure~\ref{fig:ln_pdf}) and the 300 simulations
    \citep[orange 95 per cent confidence region,][]{Cui2018}.
    Lighter shading
    indicates regions where biases due to modeling systematics may exist.
    The widely used profile shape from \citet{Nelson2014} is shown as a green line.
    The {\it Hitomi} result for Perseus is shown in pink for the range
    of observed velocity dispersions \citep{Hitomi2018}.
    Observational results from the multi-probe JACO 
    and X-COP analyses are shown
    in grey \citep[95 per confidence,][]{Siegel2018} and green \citep[$2\sigma$,][]{Eckert2019}.
    Dashed and dot-dashed lines indicate the most likely profiles for the
    more relaxed and less relaxed sub-samples of the CLASH galaxy clusters and the 300 simulations.
    The red dotted line shows the most likely profile from a CLASH-like selection
    of 20 objects from the 300 simulations.}
    \label{fig:one}
\end{figure}

\begin{figure}
  \centering
  \includegraphics[width=0.45\textwidth]{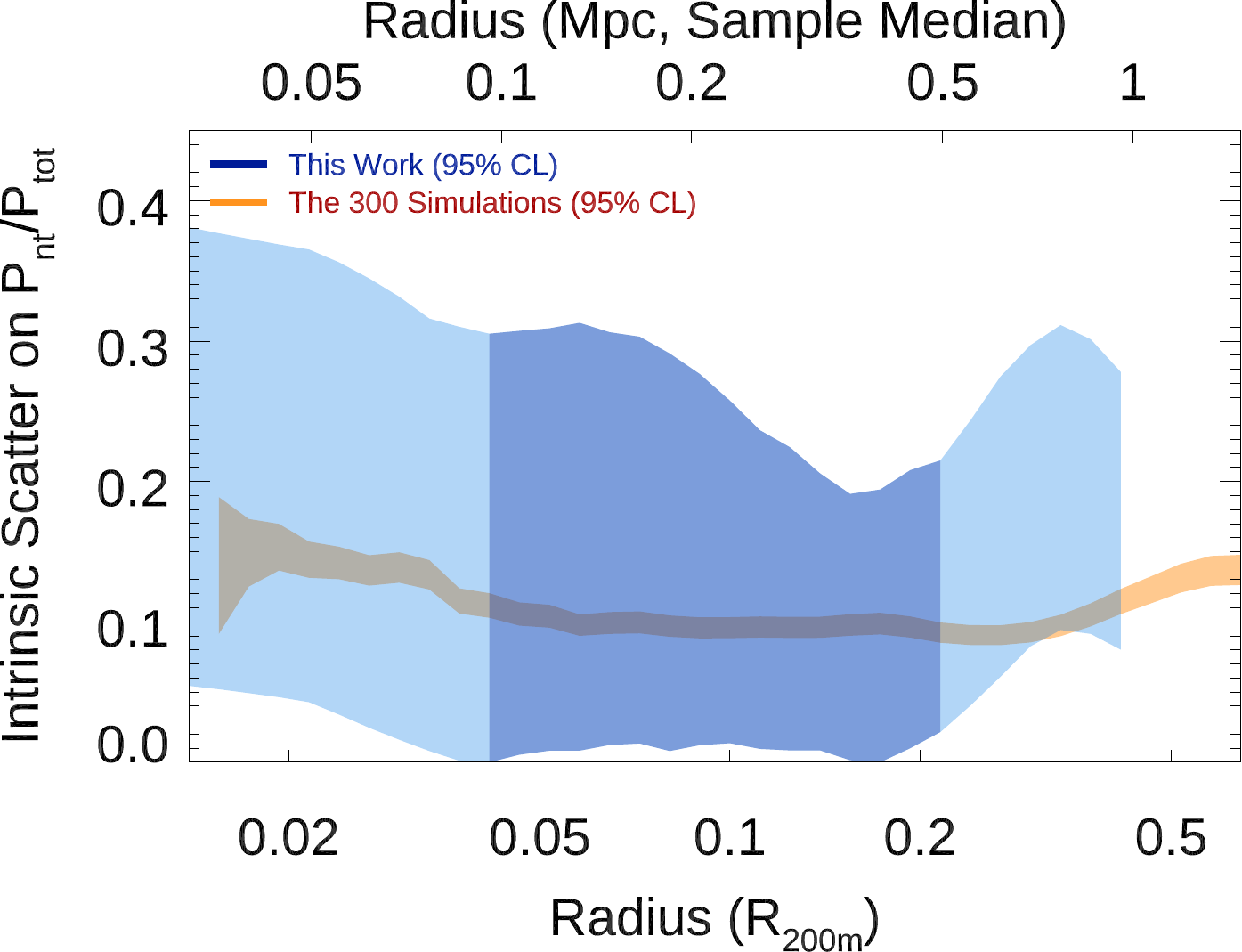}
  \caption{Intrinsic scatter on \pnt/\ptot\ (same convention as Figure~\ref{fig:one}).}
  \label{fig:scatter}
\end{figure}

Using the procedure detailed in Section~\ref{sec:methods}, we obtain 95\%
confidence regions for 
\pnt/\ptot\ profiles for all 16 individual galaxy clusters (see Figure~\ref{fig:two}),
along with the ensemble
average \pnt/\ptot\ and the intrinsic scatter about this average
(see Figures~\ref{fig:one} and \ref{fig:scatter}).
When interpreting these profiles, note
that the CLUMP-3D model assumes coalignment of the gas and total mass over
the entire radial range. While simulations indicate this assumption
is likely valid outside of 0.1\rfivec\ \citep[$\sim 0.04$\rtwom,][]{Lau2011}, 
it might not be true at smaller radii where non-gravitational processes
may play a larger role \citep{McNamara2007,Markevitch2007}.
In addition, modeling of the {\it Chandra} data is generally more
prone to systematic errors within approximately 50~kpc ($\sim 0.02$\rtwom)
due to inhomogeneities in the gas distribution \citep{McNamara2007}.
The {\it Chandra} data also set the maximum
radius that is fully constrained by GL, X-ray, and SZE data,
with \rmax\ $= 0.22$--0.68\rtwom\ (see Table~\ref{tab:sample}).
In our ensemble analysis, the maximum radius considered is 0.42\rtwom, outside
of which more than half the sample lacks {\it Chandra} coverage.

Within 0.1\rtwom, 
we measure an ensemble-average non-thermal pressure fraction
consistent with zero
(see Figure~\ref{fig:one}).
The 95 per cent confidence level upper limit takes on values in the range 8--18 per cent,
with a volume-averaged upper limit of nine per cent.
This is
consistent with the {\it Hitomi} measurements of Perseus \citep{Hitomi2018}
and the spherical JACO analysis of five galaxy clusters with extremely round
morphologies \citep{Siegel2018}.
Our analysis therefore suggests that those previous results are valid
for a larger and more diverse sample.
Recent simulations generally predict \pnt/\ptot\ values close
to our measured upper limit \citep[i.e., $\sim 10$ per cent,][]{Angelinelli2020,Gianfagna2020},
in good agreement with our analysis of the 300 simulations.

Beyond 0.1\rtwom, our measured ensemble-average \pnt/\ptot\ increases.
This is expected due to the series of shocks at varying
radii that thermalize newly accreted material \citep{Miniati2000,Molnar2012}. 
Predictions from simulations \citep{Nelson2014,Angelinelli2020,Gianfagna2020},
including our analysis of the 300 simulations, fall within
our measured 95 per cent confidence level region
at these radii. The low values of \pnt/\ptot\
obtained from the JACO and X-COP observational studies are
in mild tension with our results,
although this difference only appears at radii
where some objects require an extrapolation
beyond the {\it Chandra} data.

With a relatively large measurement uncertainty,
the CLASH data indicate an intrinsic cluster-to-cluster
scatter on \pnt/\ptot\ that is approximately constant with radius at a value
of 15--20 per cent.
This scatter is generally comparable to, or larger than, the
ensemble average value of \pnt/\ptot.
Therefore, while the gas at small radii
is highly quiescent on average, some galaxy clusters do
contain significant non-thermal pressure support within that region.
This suggests that AGN feedback may sometimes produce a larger amount
of non-thermal pressure than typical and/or there are other relevant processes
that destabilize the equilibrium near the cores of some galaxy clusters.
Similarly, while the average value of \pnt/\ptot\ increases with
radius, some galaxy clusters have very little non-thermal pressure
at large radii. This likely reflects the range of possible
accretion histories within the population \citep{Shi2014}.
Our measured scatter is consistent with what is seen in the
300 simulations and other published
simulations \citep{Nelson2014,Shi2015,Angelinelli2020,Gianfagna2020},
indicating that they accurately reproduce the
level of observed diversity.

In addition, we compute \pnt/\ptot\ profiles for sub-samples of more relaxed and
less relaxed galaxy clusters from both CLASH and the 300 simulations.
For CLASH, the equal-sized sub-samples of eight objects are chosen based
on the presence, or lack of, possible merger activity \citep{Gilmour2009,Postman2012,Mann2012}.
For the 300 simulations, a sub-sample of 101 more relaxed systems was selected
based on a centre-of-mass offset $\Delta_{\textrm{r}} \le 0.04$
and a fraction of mass in subhaloes $f_{\textrm{s}} \le 0.1$ within \rfivec\ \citep{Cui2018},
with the remaining 214 objects forming the less relaxed sub-sample.
These thresholds are commonly used, and are sufficient to identify objects
undergoing relatively minor mergers (e.g., a single 10--1 mass ratio
sub-structure will generally produce a value of $f_{\textrm{s}}$ above
our cutoff). This criteria should be similar to that used to identify
potential mergers in the CLASH objects. For instance, the lower limit
on the mass of the sub-cluster involved in the minor merger ongoing in 
MACS J1347.5+1144 is sufficient to have a mass ratio of 10--1
\citep{Johnson2012}. Thus, we expect that the more relaxed sub-sample identified
from the 300 simulation is comparable to that identified in the CLASH clusters.
However, this is less likely to be true for the less relaxed sub-sample.
As noted by \cite{Meneghetti2014}, and further indicated by the larger fraction
of less relaxed objects in the sub-sample selected from the 300 simulations
($\sim$70~per cent) compared to CLASH (50 per cent), the CLASH selection
excludes major mergers and extremely dynamically active systems
(e.g., with evidence of multiple minor mergers) which are
represented in the 300 simulations.

When considering the less relaxed and more relaxed sub-samples defined above,
it is not possible to robustly constrain the intrinsic scatter with only eight
objects in the CLASH sub-samples. Therefore, for these sub-sample fits we fix the intrinsic scatter
to the maximum likelihood value obtained from the full sample.
At all radii, the profiles of the more relaxed and less relaxed sub-samples
are consistent for both the CLASH galaxy clusters and the 300 simulations,
indicating that the ensemble average \pnt/\ptot\ is largely insensitive to
dynamical state and thus not strongly influenced by merger activity,
at least for non-major mergers.
This conclusion is consistent with the findings
of a separate analysis of the 300 simulations focused on mass
calibration \citep{Ansarifard2020},
but in contrast to what has been found in some recent
simulations \citep{Nelson2014,Gianfagna2020}.

While the average non-thermal pressure profile appears to be largely
insensitive to dynamical state, the results described above do not definitively
establish whether it is sensitive to the specific CLASH selection.
Since the CLASH objects were selected primarily based on X-ray morphology,
we follow the general approach of \cite{Meneghetti2014} to select
a CLASH-like sample of galaxy clusters from the 300 simulations.
Specifically, we consider the concentration $C$, centroid shift $w$,
axial ratio $AR$, and power ratio $P30$ determined for three
orthogonal projections of the 80 most massive objects from
the simulations (i.e., a total of 240 projections for the ``300$_{80}$'').
For each of the 16 CLASH galaxy clusters, we select
the two projections that most closely match the values for these
quantities measured by \cite{Donahue2016}. This results in a sample of
20 unique galaxy clusters from the 300 simulations with X-ray morphologies
closely matched to the CLASH galaxy clusters (see Table~\ref{tab:x-ray}).
Although there are possible hints of lower values for the average
\pnt/\ptot\ computed from this CLASH-like sub-sample compared to the full
sample, none of these differences are statistically significant given
the size of the sub-sample (see Figure~\ref{fig:one}). Furthermore, the absolute magnitude of
these differences is relatively small (i.e., $\lesssim 3$~per cent).
Thus, we conclude that the average non-thermal pressure profile
obtained from the CLASH galaxy clusters is likely to be a good
representation of the population as a whole, with no significant
selection biases.

\begin{table}
  \centering
  \caption{X-ray morphological parameters. Inter-quartile ranges are
    given for the CLASH sample \citep{Donahue2016}, a CLASH-like
    sub-sample of 20 objects selected to best match these parameters
    from the 80 most massive galaxy clusters in the 300 simulations,
    and the full set of 80 galaxy clusters.} 
  \label{tab:x-ray}
  \begin{tabular}{lccc}
    Parameter & CLASH & 300$_{80}$ (CLASH-like) & 300$_{80}$ (All) \\ \hline
    $C$ & 0.34--0.53 & 0.42--0.54 & 0.28--0.52 \\
    $w$ ($10^{-2}$) & 0.3--1.4 & 0.4--1.5 & 0.8--4.3 \\
    $AR$ & 0.87--0.95 & 0.85--0.92 & 0.72--0.87 \\
    $P30$ ($10^{-7}$) & 0.4--1.6 & 0.3--1.1 & 0.2--8.5 \\
  \end{tabular}
\end{table}

\section{Summary and Conclusions} \label{sec:summary}

From a relatively diverse sample of 16 CLASH objects,
we find that there is generally very little non-thermal pressure
support in the core regions of galaxy clusters.
Our result suggests that highly quiescent cores are not unusual
nor restricted
to a particular subset of galaxy clusters. This conclusion is further
supported by the consistency of the non-thermal pressure fraction
measured in two sub-samples of eight objects with and without
potential merger activity.
Therefore, AGN outbursts and other
relevant heating mechanisms must typically operate in a gentle manner
that preserves approximate HSE.
Outside of the core, we find that the non-thermal
pressure fraction increases with radius, as expected
due to incomplete thermalization of newly accreted material.
Furthermore, we find that the non-thermal pressure
fraction varies significantly among the population, both within
and external to the core region.
Other than the 300 simulations suggesting
a slightly higher ensemble average \pnt/\ptot\ near the galaxy clusters' centres,
we find generally good agreement between the observational results from the
CLASH sample and the results from those simulations.
At the smaller radii where indications of a possible difference do exist,
our measured upper limit is similar to the average \pnt/\ptot\ found in the 300 simulations.
Therefore, at a confidence level of approximately 95~per cent,
our result is still consistent with the simulations in this radial range.
This implies that our measurement is also consistent with the relatively small HSE-derived mass biases
obtained from the 300 simulations at those radii
\citep[e.g., the 5--10 per cent bias found at \rtwofivec\ by][]{Ansarifard2020},
suggesting that non-thermal motions in the central regions of galaxy clusters do not
significantly impact HSE-derived masses.

\section*{Acknowledgements}

The simulations  of the 300 project were performed  on  the Marenostrum Supercomputer of the Barcelona Supercomputing Center  thanks to computing time from the ``Red Espa\~nola de Supercomputaci\'on".   JS was supported by NSF/AAG award 1617022. GY was supported by MICIU/FEDER  (Spain) grant PGC2018-094975-C21. WC was supported by European Research Council grant 670193. KU was supported by the Ministry of Science and Technology of Taiwan (grants MOST 106-2628-M-001-003-MY3 and MOST 109-2112-M-001-018-MY3) and from the Academia Sinica Investigator Award (grant AS-IA-107-M01). SE, ER, and MS were supported by contract ASI-INAF n.2017-14-H.0. SE and MS were supported by INAF mainstream project 1.05.01.86.10. SE was supported by contract ASI-INAF Athena 2019-27-HH.0, and by the European Union's Horizon 2020 Programme under the AHEAD2020 project (grant 871158).

\section*{Data Availability}

The data underlying this analysis are available from repositories for CLASH {\url{https://archive.stsci.edu/prepds/clash/}}, {\it Chandra} {\url{https://cxc.cfa.harvard.edu/cda/}}, and {\it Planck} {\url{https://irsa.ipac.caltech.edu/Missions/planck.html}}.



\bibliographystyle{mnras}
\bibliography{references} 






\label{lastpage}
\end{document}